\documentstyle[12pt]{article}
\begin{document}
%
%
\def\R{\rm l\!R\,}
\def\Compton{\overline{\phantom{\overline{.\!.}}}\!\!\!\lambda}
\centerline{\bf{\Large
Bosonic Casimir effect in external magnetic field}}
\vspace{0.7cm}
\centerline{M.V. Cougo-Pinto\footnote{marcus@if.ufrj.br}, 
C. Farina\footnote{farina@if.ufrj.br}, M. R. Negr\~ao\footnote
{guida@if.ufrj.br} and A. C. Tort\footnote{tort@if.ufrj.br}}
\vspace{.5cm}
\centerline{\it  Instituto de F\'{\i}sica, Universidade Federal do Rio
de Janeiro }
\centerline{ CP 68528, Rio de Janeiro, RJ}
\centerline{ 21945-970, Brazil}

\begin{abstract}
We compute the influence of an external magnetic field on the
Casimir energy of a massive charged scalar field confined between two parallel
infinite plates. For this case the obtained result shows that the magnetic field 
inhibits the Casimir effect.
\end{abstract}
\vskip 1 true cm

The Casimir effect can be generally defined as the effect of a non-trivial
space topology on the vacuum fluctuations of relativistic quantum fields 
\cite{MostepanenkoTrunovB,PlunienMullerGreinerR}. The corresponding change 
in the vacuum fluctuations appears as a shift in the vacuum energy and an
associated vacuum pressure. This shift is known as the Casimir energy of the 
field due to the given space constraints. The original Casimir effect 
\cite{Casimir48} is the attraction of two neutral perfectly conducting parallel 
plates placed in vacuum. The boundary conditions imposed by the metallic plates
confine the vacuum fluctuations of the quantum electromagnetic field in the 
space between the plates. The effect of the boundary conditions 
can be viewed as a departure from the trivial topology of $\R^3$ to the topology of 
$\R^2\times[0,a]$, where $a$ is the distance between the plates. The resulting
shift in the vacuum energy of the quantum electromagnetic field was computed by
Casimir and is given by\cite{Casimir48}:
\begin{equation}\label{photonE}
{\cal E}_{\gamma}(a)=-\ell^2{\pi^2\over 720a^3}\; ,
\end{equation}
where $\ell^2$ is the area of each plate and the close spacing between them is 
implemented by the condition $a\ll\ell$. The pressure corresponding to 
(\ref{photonE}) was first measured by Sparnaay in 1958\cite{Sparnaay58} and more 
recently with high accuracy by Lamoreaux\cite{Lamoreaux97} and by Mohideen and 
Roy\cite{MohideenRoy98}. 

The Casimir energy has become an important ingredient of any theory with non 
trivial vacuum and has been computed for fields other than the electromagnetic one
with several types of boundary condition 
\cite{MostepanenkoTrunovB,PlunienMullerGreinerR}.
In the case of an electrically charged quantum field it poses by itself the 
question of how the charged fluctuations, and therefore the Casimir 
effect, are affected by fields coupling to the fluctuations through this charge. 
This question is strongly motivated by the fact that in a more complete picture 
of the Casimir effect the charged fluctuations of the constrained vacuum are or 
may be put under the influence of other fields. Within a hadron, for example, 
the vacuum fluctuations of quark fields are affected by the electromagnetic 
field of the quarks and by the color field of gluons and quarks. Also the vacuum 
fluctuations of gluon fields are affected by the color field of quarks and 
gluons. In this example of quarks and gluons the unavoidable influence of the 
fields on the constrained charged fluctuations is of extreme complexity. 
As a first step to the understanding of how charged fermionic and bosonic constrained
vacuum fluctuations are affected by fields coupling to this charge we may consider 
the problem in its most simple form, to wit: vacuum fluctuations of Dirac or 
scalar electrically charged fields under the influence of an external constant 
uniform magnetic field and constrained by the simplest of the possible boundary 
conditions. In the case of a constrained Dirac vacuum the external field 
enhances the Casimir energy\cite{C-PFT98}.
Here we shall consider a complex scalar field confined between 
two infinite plates with a constant uniform magnetic field in a direction 
perpendicular to the plates. The charged scalar field allows us to ignore kinematical
complexities which are not relevant to an initial approach to this problem.
The choice of a pure magnetic field excludes the possibility 
of pair creation for any field strength. The confinement between infinite plates 
is described by a simple form of Dirichlet boundary condition and the direction 
of the magnetic field perpendicular to the plates is obviously a simplifying 
choice.  Under such assumptions the formalism may be kept simple in order for 
us to concentrate on the fundamental issue, which is the physical effect of 
the external field on the Casimir effect. Once the main feature of such influence 
is determined the path is open to consider more complicated geometries and 
external fields as well as other quantum vacua. Notice that we consider a problem
in which the charged quantum vacuum is constrained by the boundary conditions and
the external electromagnetic field is not. Therefore the influence of the external 
field on the charged vacuum already appears at the one-loop level, at which our
calculations will be done. In contrast we have 
in the Scharnhorst effect 
\cite{Scharnhorst90,Barton90,BartonScharnhorst93,C-PFST98a,C-PFST98b} 
that boundary conditions on a pair of parallel plates are imposed on 
the electromagnetic quantum vacuum but not on the charged vacuum of electrons 
and positrons. As a result there is a change in the velocity of propagation 
of an external electromagnetic
wave in the region between the plates. The Scharnhorst effect involves 
two-loop diagrams because the coupling between the external field and the 
quantum electromagnetic field requires the intermediation of a charged fermion 
loop.

Let us calculate the Casimir energy of the charged scalar field in a constant
aplied magnetic field using a method introduced by Schwinger to obtain the Casimir 
energy \cite{Schwinger92} from the proper-time representation of the effective 
action\cite{Schwinger51}. Since the method has been clearly explained by Schwinger 
\cite{Schwinger92} and already applied to several situations
\cite{C-PFSST} we may use it here without going into too much 
detail. We start with Schwinger's proper-time formula for the effective action 
\cite{Schwinger51}:
\begin{eqnarray}
{\cal W} = - \frac{i}{2} \int_{s_0}^{\infty}
\frac{ds}{s} Tr e^{-isH},
\label{acao}
\end{eqnarray}
where $s_{0}$ is a cutoff in the proper-time $s$, $Tr$ means the total
trace and $H$ is the proper-time Hamiltonian, which is given by $(p -
eA)^{2} + m^{2}$, where $p_{\mu} = - i {\partial}_{\mu}$, $e$ is the charge
of the scalar field, $A$ is the electromagnetic potential and $m$ is
the mass of the scalar field. The boundary condition gives for the component of
the momentum which is perpendicular to the plates the eigenvalues
$n\pi/a$,where $n$ is a positive
integer. The other spatial components of the momentum are constrained into
the Landau levels created by the magnetic field ${\bf B}$ and we choose the
direction of ${\bf B}$ in such a way that $eB$ is positive. The trace in
(\ref{acao}) is given by:
\begin{eqnarray}
Tr e^{-isH} = 2 e^{-ism^{2}}\sum_{n'=0}^\infty
\frac{{\ell}^{2} eB}{2\pi}
e^{-iseB(2n'+1)} \sum_{n=1}^\infty e^{-is(n\pi
/a)^{2}} \int \frac{dtd\omega}{2\pi}
e^{is\omega^{2}}
\label{traco1}
\end{eqnarray}
where the factor $2$ is due to the two degrees of freedon in the complex field;
the first sum is over the Landau levels with the corresponding multiplicity
factor due to degeneracy; the second sum is over the eigenvalues stemming
from the Dirichlet boundary conditions and the integral range is given by
the measurement time $T$ and by the continuum of eigenvalues $\omega$ of
the operator $p^{0}$. Following Schwinger's regularization prescription
\cite{Schwinger92} we apply Poisson sum formula
\cite{Poisson1823} to the second sum in order to
obtain:
\begin{eqnarray}
\sum_{n=1}^\infty e^{-is(n\pi/a)^{2}} =
\frac{a}{\sqrt{i\pi s}}
\sum_{n=1}^\infty e^{i(an)^{2}/s} +
\frac{a}{2\sqrt{i\pi s}} - \frac{1}{2}.
\label{poisum}
\end{eqnarray}
The sum over the Landau levels is straightforward and leads
to:
\begin{eqnarray}
\sum_{n'=0}^\infty \frac{eB{\ell}^{2}}{2\pi}
e^{-iseB(2n'+1)} = \frac{eB{\ell}^{2}}{4\pi}
cosech(iseB).
\label{somaland}
\end{eqnarray}
Using (\ref{poisum}) and (\ref{somaland}) into (\ref{traco1}), we obtain for 
the trace:
\begin{eqnarray}
Tr e^{-isH} = \frac{a{\ell}^{2}T}{4{\pi}^{2}}\frac{e^{-ism^{2}}}{is^{2}}[1 +
iseB{\cal M}(iseB)] \biggl [ \frac{1}{2} - \frac{\sqrt{i\pi s}}{2a} +
\sum_{n=1}^\infty e^{i(an)^{2}/s} \biggl ],
\label{traco2}
\end{eqnarray}
where ${\cal M}$ is the function defined by:
\begin{eqnarray}
{\cal M}(\xi) = cosech{\xi} - {\xi}^{-1}.
\label{myfunc}
\end{eqnarray}
Substituting now equation (\ref{traco2}) into equation (\ref{acao}) we get the
effective action:
\begin{eqnarray}
{\cal W} = {\cal L}^{(1)}(B)Ta{\ell}^{2} -
{\cal E}(a,B)T,
\label{acao2}
\end{eqnarray}
where on the right hand side the first term gives the (unrenormalized) effective
Lagrangian:
\begin{eqnarray}
{\cal L}^{(1)}(B) = -\frac{1}{16{\pi}^{2}}
{\int_{s_{0}}}^{\infty}
\frac{ds}{s^{3}} e^{-ism^{2}} (iseB)
cosech(iseB)
\label{lag}
\end{eqnarray}
and the second term gives the (still cutoff-dependent) Casimir energy:
\begin{eqnarray}
{\cal E}(a,B) =
-\frac{a{\ell}^{2}}{8{\pi}^{2}} \sum_{n=1}^\infty
{\int_{s_{0}}}^{\infty}
\frac{ds}{s^{3}} e^{-ism^{2} + i(an)^{2}/s} [ 1+ iseB{\cal
M}(iseB)],
\label{encas1}
\end{eqnarray}
which is the quantity we are interested in. The effective Lagrangian
${\cal L}^{(1)}(B)$  is analogous to the Euler-Heisenberg Lagrangian for
the fermionic case \cite{H-E} and was first obtained by Schwinger 
in 1951 \cite{Schwinger51}.
Since it does not depend on $a$ it makes no contribution to the Casimir energy.
Usually, spurious terms must be subtracted before eliminating the cutoff $s_o$ in
(\ref{encas1}) but in the present calculation they were all left in the effective 
Lagrangian which is of no concern to us here. So we may simply take $s_o=0$ in 
(\ref{encas1}). Continuing with Schwinger's method we now use Cauchy theorem to
make a $\pi/2$ clockwise rotation of the integration $s$-axis, which results in 
a substitution of $s$ by $-is$ in the integrand of (\ref{encas1}). Part of 
this integrand can be expressed in terms of the modified
Bessel function $K_{2}$ (cf. formula {\bf 3471.9} in 
\cite{GradshteijnRyzhik65}) and (\ref{encas1}) reduces to:
\begin{eqnarray}
\frac{{\cal E}(a,B)}{{\ell}^{2}} &=& -
\frac{(am)^{2}}{4{\pi}^{2}a^{3}} \sum_{n\in N}\frac{1}{n^{2}}
K_{2}(2amn) + \nonumber\\
&-& \frac{1}{8{\pi}^{2}a^{3}} \sum_{n=1}^\infty
{\int_{0}}^{\infty} \frac{ds}{s^3} e^{-s(am)^{2}-n^{2}/s} 
s\,eBa^2{\cal M}(s\,eBa^{2}).
\label{encas2}
\end{eqnarray}
The first term on the right hand side of this equation is the usual Casimir energy 
in the absence of the external magnetic field:
\begin{eqnarray}
\frac{{\cal E}(a,0)}{a{\ell}^{2}} = - \frac{(am)^{2}}{4{\pi}^{2}a^{4}}
\sum_{n=0}^\infty \frac{1}{n^{2}} K_{2}(2amn),
\label{encas0}
\end{eqnarray}
which is a result already known in current
literature\cite{MostepanenkoTrunovB,PlunienMullerGreinerR}; in the limit
$m\rightarrow 0$ this result reduces to (\ref{photonE})
(actually because the complex scalar field
has two degrees of freedon and the photon field two polarizations). Here we are 
interested in the second term on the right hand side of equation (\ref{encas2}):
\begin{eqnarray}
\frac{{\Delta\cal E}(a,B)}{{\ell}^{2}} =
- \frac{1}{8{\pi}^{2}a^{3}} \sum_{n=1}^\infty
{\int_{0}}^{\infty} \frac{ds}{s^3} e^{-s(am)^{2}-n^{2}/s} 
s\,eBa^2{\cal M}(s\,eBa^{2}).
\label{Delta}
\end{eqnarray}
which measures the influence of
the external magnetic field in the Casimir energy. Due to the simple behaviour of the
function ${\cal M}$ defined in (\ref{myfunc}) we can determine the main features of 
this influence. The function $-\xi\,{\cal M}(\xi)$ increases monotonically from zero 
to the asymptotic value $1$ when $\xi$ goes from $0$ to $\infty$. Therefore we see 
that the external magnetic field always inhibits the Casimir energy of the scalar 
field and suppress it completely in the limit $B\rightarrow\infty$. This is the 
result that answer the question raised above. This result should be contrasted with 
the result for a Dirac field for which the Casimir energy is always enhanced by the 
external magnetic field \cite{C-PFT98}. It is very interesting that fermionic 
and bosonic charged 
vacuum present such a clear and opposite behaviours in presence of an external 
magnetic field. We didn't find an intuitive explanation for those behaviours but 
it is quite possible that it is related with the paramagnetic and diamagnetic
characters of fermionic and bosonic vacua, respectively. At any rate it is 
important to take in consideration this opposite behaviour of bosonic and fermionic 
vacua in the presence of an external field, because these vacua actually exist
toghether in the presence of fields and may also be constrained by boundary
conditions, as remarked above. 
Let us notice, for example, that the shift in the zero point energy caused by the 
external field depends on the mass of the field in the bosonic case (\ref{Delta})
as well as in the fermionic case \cite{C-PFT98}, and therefore a cancellation of 
zero point energies of those vacua depends on the specific relations between the 
masses of the quantum fields.

It is also instructive to define
\begin{equation}\label{m_B}
m_B=\sqrt{m^2+eB}
\end{equation}
and write the complete Casimir energy (\ref{encas2}) as:
\begin{eqnarray}
\frac{{\cal E}(a,B)}{{\ell}^{2}} =
- \frac{1}{8{\pi}^{2}a^{3}} \sum_{n=1}^\infty
{\int_{0}}^{\infty} ds\,s^{-3} e^{-s(am_B)^{2}-n^{2}/s} 
\frac{2s\,eBa^2}{1-e^{-2s\,eBa^2}}
\label{E(m_B)}
\end{eqnarray}
Comparing this expression with its limit when $B\rightarrow 0$ we may say that 
the effect of the external magnetic field on the usual Casimir energy is given
in the integrand of (\ref{E(m_B)}) by the $B$ dependent fraction  and the constant 
$m_B$ which appears in the exponential. When $B\rightarrow 0$ the fraction tends to 1
and $m_B\rightarrow m$. For a strong magnetic field the exponential is the dominant
factor in the integrand and the effect of the magnetic field on the Casimir energy
appears roughly as the substitution of $m$ by $m_B$; certainly the $B$ dependent 
fraction may still in this case affect the precise influence of the magnetic field 
on the Casimir energy.

Let us consider the strong field regime, in which changes in the charged
vacuum should be more prominent. The integral in equation (\ref{encas2}) is
dominated by the exponential function whose maximum is $e^{-2amn}$ and
occurs at $\sigma = am/n$. Due to this feature we are justified in
substituting the function ${\cal M}(\xi)$ by $2e^{-\xi} - {\xi}^{-1}$ if
$B>>({\phi}_{0}/a^{2})(a/{\lambda}_{c})$, where ${\phi}_{0}$  is the
fundamental flux $1/e$ and ${\lambda}_{c}$ is the Compton wavelength $1/m$.
Therefore, in the strong field regime, the second term in (\ref{encas2}) can
also be expressed in terms of a modified Bessel function (formula {\bf
3471,9} in ref. \cite{GradshteijnRyzhik65}), and the Casimir energy can be written
as:
\begin{eqnarray}
\frac{{\cal E}(a,B)}{{\ell}^{2}} = - \frac{eBa^2}{2{\pi}^{2}a^{3}} \sqrt{(am)^{2}
+ eBa^{2}} \sum_{n\in N} \frac{1}{n} K_{1}(2n\sqrt{(am)^{2} +
eBa^{2}}).
\label{encasf}
\end{eqnarray}
Notice that the sign in the square root is to be expected because in the regime
we are working with a minus sign means energy creation or anihilation
which cannot happen when we are dealing with a constant and uniform magnetic field.
We can also use (\ref{m_B}) to rewrite (\ref{encasf}) in the following form:
\begin{eqnarray}
\frac{{\cal E}(a,B)}{{\ell}^{2}} = - eBa^2\frac{am_B}{2{\pi}^{2}a^{3}}
\sum_{n=1}^\infty \frac{1}{n} K_{1}(2am_B n)\; ,
\label{encasf(m_B)}
\end{eqnarray}
which is in a more appropriated form to compare with (\ref{encas0}). 
By further stressing the 
strong field regime we can take the asymptotic limit of $K_1$ (cf. {\bf 8446} in 
\cite{GradshteijnRyzhik65}) in (\ref{encasf(m_B)}) with $m_B\approx eB$ to obtain:
\begin{equation}
\frac{{\cal E}(a,B)}{{\ell}^{2}} = \frac{(eBa^2)^{5/4}}{a^3}e^{-2\sqrt{eBa^2}}\, .
\end{equation}
Turning now our attention to the weak field regime $B<<({\phi}_{0}/a^{2})(a/{\lambda}_{c})$
we can substitute in the integrand of (\ref{Delta}) 
$\xi{\cal M}(\xi)$ by $-\xi^2/6$ to obtain:
\begin{equation}
\frac{\Delta{\cal E}(a,B)}{{\ell}^{2}} = -\frac{(eBa^2)^2}{24{\pi}^{2}a^{3}}
\sum_{n=1}^\infty K_{0}(2am n)\; ,
\label{encasfr}
\end{equation}
Summarizing the results we have in equation (\ref{encas2}) the exact expression
for the influence of the external magnetic field on the Casimir energy of a scalar 
charged field. Equations (\ref{encasf}) and (\ref{encasfr}) particularize the 
result of equation (\ref{encas2}) to the regimes of strong and weak magnetic field,
respectively. In any case the external  field inhibits the Casimir energy of the
scalar field. This is in opposition with the case of a Dirac field, whose
Casimir energy is enhanced by the external magnetic field \cite{C-PFT98}.
We may also look at the interplay between constraints
and external field on the quantum vacuum from a completely different point of view. 
Instead of asking what is the influence of the magnetic field on the Casimir 
energy of the constrained vacuum we can ask what is the effect that constraints 
on the vacuum have on the effective Lagrangian 
for the magneticf field. This study has already 
been done for the fermionic vacuum \cite{C-PFRT98} and will in the near future 
be presented also for the bosonic vacuum. It would also be interesting to 
investigate the effect
of an external magnetic field on the bosonic vacuum of a scalar field with
space-time symmetry given by the $\kappa$-deformed Poincar\'e algebra
\cite{C-PF97} in order to see the relation between the inhibiting effect of 
the magnetic field on the Casimir energy and the mechanism of creation of 
field excitations due to the deformation.

\end{document}